\documentclass[10pt, conference, compsocconf]{IEEEtran}
\ifCLASSINFOpdf
\else
\fi
\usepackage{xcolor}
\usepackage{graphicx}
\usepackage{caption}
\usepackage{subcaption}
\graphicspath{ {img/} }
\usepackage{tikz}
\usepackage{listings}
\usepackage{url}

\newcommand{\tikzcircle}[2][red,fill=red]{\tikz[baseline=-0.5ex]\draw[#1,radius=#2] (0,0) circle ;}%https://tex.stackexchange.com/questions/89821/how-to-draw-a-solid-colored-circle

\newcommand{\yes}{\tikzcircle[blue, fill=blue]{4pt}}
\newcommand{\some}{\tikzcircle[blue, fill=white]{4pt}}
\newcommand{\no}{\tikzcircle[white, fill=white]{4pt}}

\begin{document}
\title{MORF: A Framework for Predictive Modeling and Replication At Scale With Privacy-Restricted MOOC Data}


% \author{\IEEEauthorblockN{Anonymized For Review}
% \IEEEauthorblockA{line 1 (of Affiliation): dept. name of organization\\
% line 2: name of organization, acronyms acceptable\\
% line 3: City, Country\\
% line 4: Email: name@xyz.com}
% \and
% \IEEEauthorblockN{Anonymized For Review}
% \IEEEauthorblockA{line 1 (of Affiliation): dept. name of organization\\
% line 2: name of organization, acronyms acceptable\\
% line 3: City, Country\\
% line 4: Email: name@xyz.com}
% }

\author{\IEEEauthorblockN{Josh Gardner, Christopher Brooks}
\IEEEauthorblockA{School of Information\\
The University of Michigan\\
Ann Arbor, USA\\
\{jpgard, brooksch\}@umich.edu}
\and
\IEEEauthorblockN{Juan Miguel Andres, Ryan S. Baker}
\IEEEauthorblockA{Graduate School of Education\\
The University of Pennsylvania\\
Philadelphia, USA\\
miglimjapandres@gmail.com, rybaker@upenn.edu}
}


\maketitle

\begin{abstract}
% The abstract goes here. DO NOT USE SPECIAL CHARACTERS, SYMBOLS, OR MATH IN YOUR TITLE OR ABSTRACT.
Big data repositories from online learning platforms such as Massive Open Online Courses (MOOCs) represent an unprecedented opportunity to advance research on education at scale and impact a global population of learners. To date, such research has been hindered by poor reproducibility and a lack of replication, largely due to three types of barriers: experimental, inferential, and data. We present a novel system for large-scale computational research, the MOOC Replication Framework (MORF), to jointly address these barriers. We discuss MORF's architecture, an open-source platform-as-a-service (PaaS) which includes a simple, flexible software API providing for multiple modes of research (predictive modeling or production rule analysis) integrated with a high-performance computing environment. All experiments conducted on MORF use executable Docker containers which ensure complete reproducibility while allowing for the use of any software or language which can be installed in the linux-based Docker container. Each experimental artifact is assigned a DOI and made publicly available. MORF has the potential to accelerate and democratize research on its massive data repository, which currently includes over 200 MOOCs, as demonstrated by initial research conducted on the platform. We also highlight ways in which MORF represents a solution template to a more general class of problems faced by computational researchers in other domains.

\end{abstract}

\begin{IEEEkeywords}
data infrastructure; education; MOOC; reproducibility; predictive modeling; machine learning

\end{IEEEkeywords}

% \fi
\IEEEpeerreviewmaketitle

\section{Introduction}\label{sec:introduction}

% This is a \tikzcircle{4pt}, \tikzcircle[blue, fill=white]{4pt} followed by \tikzcircle[green, fill=blue]{1.5pt}.

\subsection{Educational Big Data in the MOOC Era}

Since the launch of the three major Massive Open Online Course (MOOC) platforms in 2012, MOOCs have been met variously with enthusiasm, curiosity, and criticism regarding their potential to transform the educational landscape and provide unprecedented access to high-quality educational content and credentials for learners around the globe. The massive, granular, and multimodal data generated by MOOCs has the potential to answer many of the research questions underlying these perspectives, informing educational research on the process of learning within online environments. To date, MOOC research has addressed a diverse array of research questions from psychometrics and social psychology to predictive modeling and machine learning. For example, several works have explored prediction of various student outcomes using behavioral, linguistic, and assignment data from MOOCs to evaluate and predict various student outcomes including course completion \cite{Kloft2014-kb, Veeramachaneni2014-ug, Yang2013-zq}, assignment grades \cite{Kizilcec2015-gu}, Correct on First Attempt
(CFA) submissions \cite{Brinton2015-ya}, student confusion \cite{Yang2015-gy}, and changes in behavior over time \cite{Bote-Lorenzo2017-yh}. A key area of research has been methods for feature engineering, or extracting structured information from raw data (i.e. clickstream server logs, natural language in discussion posts) \cite{Gardner2018-lp}.

\subsection{A Confluence of Challenges for MOOC Research}

The challenges of doing big data research have grown over the past decade as the data, statistical models, and technical tools have become increasingly complex, and MOOC research has not been immune to the many challenges facing the larger big data research community. Issues with reproducibility and the lack of replication studies have the potential to hamper the field and limit researchers' understanding, and, consequently, to diminish the potential impact of MOOCs for learners around the globe. Strict privacy regulations also limit the sharing of learner data for MOOC research. Many universities interpret the Family Educational Rights and Privacy Act (FERPA), the IRB Common Rule, or other data regulations in ways that severely limit access to MOOC data. The recent passage of the General Data Protection Regulation (GDPR) in the European Union may further limit data sharing and collection.

\subsection{Contribution: A Proposed Solution}

In this work, we present a novel big data research platform, the MOOC Replication Framework (MORF), which is designed to jointly address technical, data, and methodological barriers to reproducible and replication research in MOOCs. We present empirical evidence on the current state of the ``replication crisis'' in big data and the learning sciences in Section \ref{sec:replication-crisis}, motivating the need for MORF. Then, we provide a detailed overview of the MORF platform in Section \ref{sec:morf}, detailing MORF's unique platform-as-a-service (PaaS) architecture for working with massive privacy-restricted MOOC datasets in two modes of analysis (predictive modeling and production rule analysis). In particular, we highlight its use of containerization to ensure full replicability of computational results, and its massive, highly diverse datasets available for analysis in a secure, high-performance environment. We discuss how MORF represents a framework which can apply broadly to many other research domains, the software's openness and reuse potential, and the many benefits of big data replication research, in Section \ref{sec:discussion}.

\section{The Replication Crisis: An Empirical Perspective}\label{sec:replication-crisis}

An emerging body of empirical evidence is revealing the extent to which several factors -- a lack of basic reproducibility practices in big data research; a dearth of replication studies; and experimental, data, and inferential challenges to big data research -- have hindered the field.

\subsection{Reproducibility in Big Data Modeling Research}\label{sec:reproducibility-big-data}

Recent evidence has demonstrated that issues with basic reproducibility -- the ability to regenerate the original results of an experiment using the original methods and data -- are widespread in the field of big data research, which covers a variety of methods spanning artificial intelligence, machine learning, data mining, and simulation-based research. In a survey of 400 papers from leading artificial intelligence venues, none documented all aspects necessary to fully reproduce the work; on average, only 20-30\% of the factors needed to replicate the original work were reported \cite{Gundersen2017-zl}. In a survey of 30 text mining studies, none of the 30 works surveyed provided source code, and only one of 16 applicable studies provided an executable, to reproduce their experiments \cite{Olorisade2017-wv}; lack of access to data, software environment, and implementation methods were noted as barriers to reproducibility in the works surveyed. Even when code is available, it is often insufficient to reproduce an experiment: in a survey of 613 published computer systems papers, the published code accompanying failed to build or run in 20\% of cases; in total, it was not possible to verify or reproduce 75.1\% of studies surveyed using the artifacts provided in publication  \cite{Collberg2014-oh}.

%In a case study in reinforcement learning (RL), \cite{Henderson2018-qd} show that the variance inherent to statistical algorithms, the use of different hyperparameter settings, and even different random number generation seeds have a direct impact on whether experimental results and baseline model implementations replicate in simulation-based RL experiments. 

Much-needed replication studies -- where the original methods of an experiment are applied to \textit{new} data to evaluate the generalizability of findings and contextualize the results -- are not even possible with a great deal of published big data research, as even mere reproducibility is not possible.

\subsection{Lack of Replication Studies in Learning Sciences}

Replication research in the domain of the learning sciences has been scarce to date. A survey of the top 100 education journals \cite{Makel2014-kd} demonstrated that only 0.13\% of education articles were replications (a replication rate eight times lower than in the field of psychology). 67.4\% of replications attempted replicated the original results fully, 19.5\% replicated some (but not all) findings, and 13.1\% failed to replicate any original findings.

Replication specific to MOOCs and educational big data research has been particularly scarce. This may make the existing body of research in MOOCs particularly unreliable, an issue compounded by the fact that most MOOC studies to date have used small samples of MOOCs, and selected highly-varying subsets of students from the available datasets (e.g. only students who joined in the first 10 days of the course and have viewed at least one video; completed pre-course survey and the first end-of-unit exam, etc.) \cite{Gardner2018-lp}. In other domains, it has been shown that selecting subpopulations which vary in even subtle ways has led to very different experimental conclusions in big data analyses depending on the population used \cite{Johnson2017-uq}.

The limited MOOC replication research to date has shown that many published findings in the field may not replicate. For example, in a large-scale replication, \cite{Andres2018-mt} found that only 12 of 15 previously-published MOOC findings replicated significantly across the data sets, and that two findings replicated significantly in the opposite direction; similar results in a machine learning replication in the context of algorithm and feature selection were shown in \cite{Gardner2018-wo}. 

% In an attempt to replicate the findings of the ``deep knowledge tracing'' method introduced in \cite{Piech2015-yy}, \cite{Khajah2016-md} demonstrated that much simpler baseline methods could achieve equivalent performance, and that the initial performance gains demonstrated in the original work were at least partially due to data leakage (this work was not in a MOOC, but demonstrates potential replicability issues with computational big data research in the field of education).

\subsection{Three Barriers to Reproducible Big Data Research}\label{sec:barriers}

We identify three key sets of barriers contributing to both the lack of general reproducibility in big data research, and the lack of replication studies within the field of educational big data research in particular. MORF attempts to resolve all three barriers. 

\textbf{Experimental challenges} with reproducibility relate to reproducing the exact experimental protocol  \cite{Gundersen2017-zl}. Many have advocated for the open sharing of code as a potential solution to address technical issues with reproducibility \cite{Stodden2013-bh}. However, as discussed in Section \ref{sec:reproducibility-big-data}, even when code is available, other technical issues can prevent reproducibility in computational research workflows \cite{Donoho2015-aq, Kitzes2017-pf}.

Researchers have argued for over two decades that the complete software environment is a necessary condition for reproducing computational results \cite{Buckheit1995-zw}; however, the open sharing of such environments remains rare. Even when code is bug-free, compiles correctly, and is publicly shared, issues that are not resolved by code-sharing include (i) \textit{code rot}, in which code becomes non-functional or its functionality changes as the underlying dependencies change over time (for example, an update to a data processing library which breaks backwards compatibility, or a modified implementation of an algorithm which changes experimental results), as well as (ii) \textit{dependency hell}, in which configuring the software dependencies necessary to install or run code prevents successful execution \cite{Boettiger2015-cu}. This complex web of interdependencies is rarely described or documented in published machine learning and computational science work \cite{Donoho2015-aq, Gundersen2017-zl, Olorisade2017-wv}.

\textbf{Methodological challenges} to reproducibility reflect the methods of the study, such as its procedure for model tuning or statistical evaluation. Existing work on reproducibility focuses on technical challenges, but methodological issues are at least as important. Methodological challenges include the use of biased model evaluation procedures \cite{Cawley2010-la, Varma2006-by}, the use of improperly-calibrated statistical tests for classifier comparison \cite{Dietterich1998-vh}, or ``large-scale hypothesis testing'' where thousands of hypotheses or models are tested at once, despite the fact that most multiple testing corrections are not appropriate for such tasks \cite{Efron2016-to}. A machine learning version of these errors is seen in massive unreported searches of the hyperparameter space, and in ``random seed hacking'' wherein the random number generator itself is systematically searched in order to make a target model's performance appear best or a baseline model worse \cite{Henderson2018-qd}. Replication platforms can address methodological issues -- working toward what \cite{Gundersen2017-zl} terms \textit{inferential reproducibility}-- by architecting platforms which support effective methodologies and adopt them by default, effectively nudging researchers to make sound choices. 

\textbf{Data challenges} to reproducibility concern the availability of data itself. In some domains of big data research, making raw data available is more an issue of convention than a true barrier to reproducibility. However, educational data are governed by strict privacy regulations which protect the privacy of student records. Similar restrictions affect other big data domains, from the health sciences to computational nuclear physics \cite{Kitzes2017-pf}. As a result, researchers are often legally prohibited from making their data available. Efforts such as the Pittsburgh Science of Learning Center DataShop \cite{Koedinger2010-yg} and the HarvardX MOOC data sets \cite{HarvardX2014-ds} have attempted to address this problem in educational research by only releasing limited non-reidentiﬁable data, but many analyses require the original, unprocessed data for a full replication. Restricted data sharing is one of the main factors (in our experience) hindering generalizability analysis in educational data mining: investigators are generally limited to one or two courses worth of data (e.g. the courses they instruct), and models are often overfit to these datasets.

\section{The MOOC Replication Framework (MORF)}\label{sec:morf}

\begin{figure*}
    \centering
    \includegraphics[width = 0.8 \textwidth]{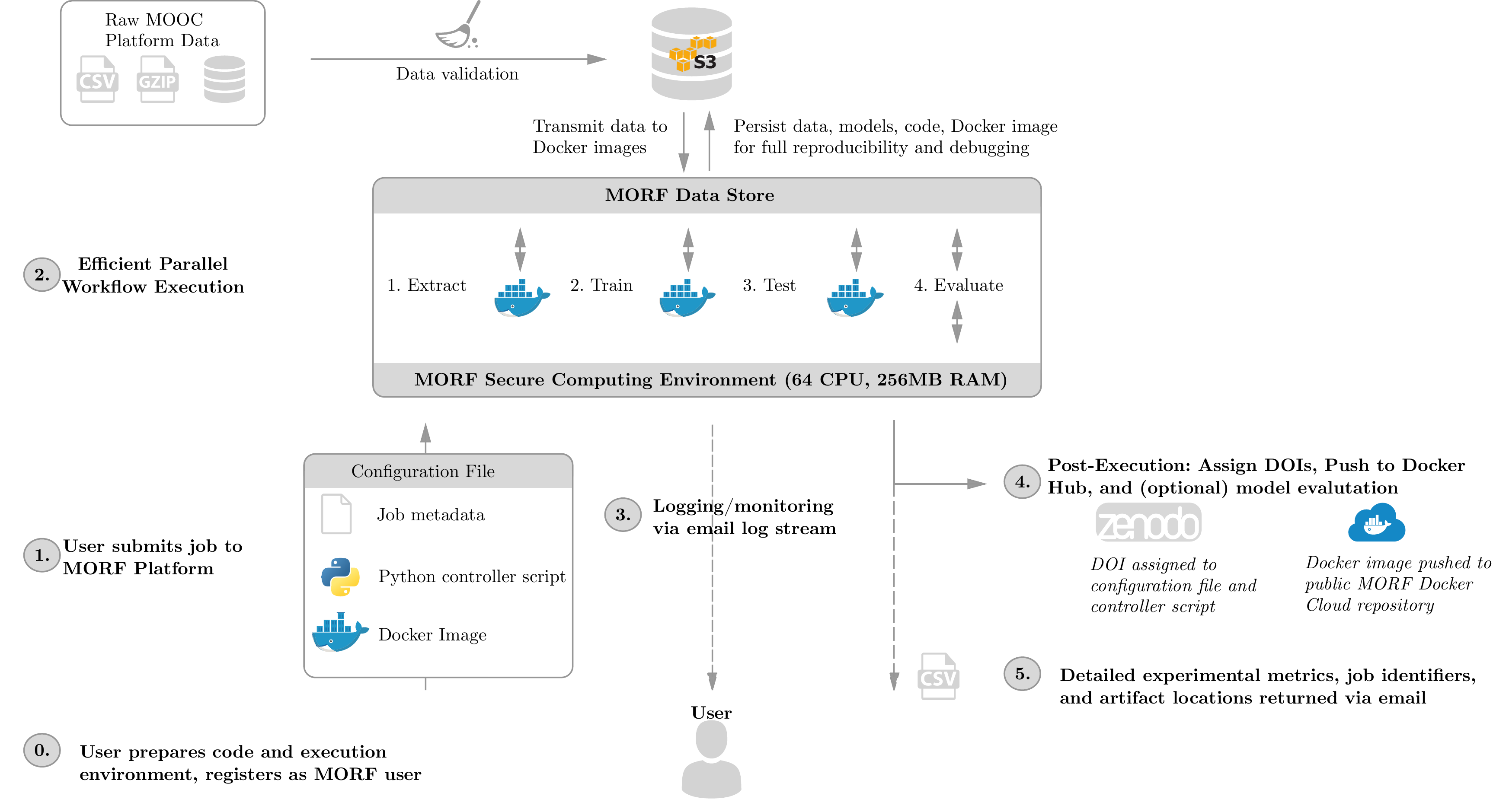}
    \caption{MORF platform architecture: this figure shows a high-level overview of a predictive modeling workflow in MORF. Solid lines indicate data transmission/archiving for reproducibility; dashed lines indicate messaging to MORF users.}
    \label{fig:morf-architecture}
\end{figure*}

MORF's key design principles are: (i) ensure the complete end-to-end reproducibility of experiments, (ii) support high-quality replication studies and original research using multiple methods of analysis, (iii) provide access to a large and diverse dataset, (iv) leverage high-performance computing requiring minimal user input, and (v) ensure complete compliance with legal restrictions on data sharing. This section detail's MORF's architecture and how it encodes these principles. In Section \ref{sec:discussion}, we discuss how this architecture may apply to a much broader set of problems across many other domains.

\subsection{Platform Architecture}

MORF consists of two main components: an open-source Python API for specifying the workflow of an experiment (the ``MORF API''), and a Platform-as-a-Service (PaaS), which is a running instance of MORF's back-end infrastructure coupled with computational resources and a large MOOC dataset (the ``MORF platform'').

\subsubsection{MORF API and Controller Scripts}

The life-cycle of a complete end-to-end experiment, from raw data to results, is shown in Figure \ref{fig:morf-architecture}. First, a user creates and submits a configuration file to MORF, either using an HTTP request or using the \texttt{easy\_submit()} MORF API function. This configuration file contains job metadata, including a pointer to an executable Docker image which encapsulates all code, software, and operating system dependencies for the users' experiment. The configuration file also points to a Python controller script that specifies the high-level experimental workflow, such as how model training and testing should occur and whether cross-validation or a holdout set should be used in a predictive modeling experiment. The use of controller scripts is a best practice for reproducible computational research \cite{Kitzes2017-pf}, as it provides a single script to fully reproduce an experiment. An additional advantage is that MORF controller scripts are human-readable, providing a high-level overview of an experiment. An example of a controller script from the experiment in \cite{Gardner2018-ad} is shown in Listing 1.

MORF uses the controller script to manage low-level data platform tasks, including (i) data wrangling (retrieving and archiving necessary data at each step of the experiment); (ii) Docker image setup and execution; and (iii) parallelization. Of particular note is (iii), as MORF is able to leverage parallelization without requiring any user input or code written for parallel execution. The controller script provides sufficient information about how MORF can execute parallelization, which can lead to speedups of 1-2 orders of magnitude when each of MORF's CPUs is occupied with a separate task (e.g. training models on each of the different MOOC courses available).

\begin{lstlisting}[language=Python, caption=A MORF controller script, linewidth = 0.95\columnwidth]
extract_session()
extract_holdout_session()
train_course(label_type = `dropout')
test_course(label_type = `dropout')
evaluate_course(label_type = `dropout')
\end{lstlisting}

\subsubsection{Docker Containerization for Reproducibility}

Another key component of MORF's architecture is that it requires submissions as executable Docker containers in order to resolve many of the experimental challenges described in Section \ref{sec:barriers}. Docker containers can be thought of as lightweight virtual machines which fully encapsulate the code, software dependencies, and execution environment of an end-to-end experiment in a single file.

Docker containers were originally developed to resolve the technical issues described above in software development contexts \cite{Merkel2014-xu}, and are frequently used in both industrial software applications as well as computational and computer systems research \cite{Boettiger2015-cu, Cito2016-ej, Jacobsen2015-bq}. Their use in data science applications is increasing, but the execution, publication, and sharing of pre-built Docker images as part of a research workflow is rare. While some existing platforms utilize containerization (e.g. Codalab\footnote{\url{http://codalab.org/}}), this functionality is hidden from the user and containers are created ``under the hood'' by the platform. This limits users' ability to fully leverage containerization by building the complex, customized environments many machine learning experiments may require or to share these containers upon completion of an experiment. We are not aware of any data research platform which allow users to submit Docker images directly for execution.

A major advantage of Docker over simple code-sharing is that Docker containers fully reproduce the entire execution environment of the experiment, including code, software dependencies, and operating system libraries, exactly as this environment is configured at the time of an experiment. These containers are much more lightweight than a full virtual machine, but achieve the same level of reproducibility \cite{Merkel2014-xu, Jacobsen2015-bq}. 

When submitting a job for execution to the MORF platform, a user generates a Docker image containing the code, software, and operating system dependencies required to execute their experiment, and uploads the image to a public location (files located locally, HTTP, or in Amazon S3 are supported). The user provides the image's URL the configuration file submitted to MORF, and the image is fetched, checked, and executed according to the controller script. When an experiment completes error-free execution, MORF uploads the image to a public image repository on Docker Hub using a unique identifier. This makes implementations of every experiment on MORF immediately and publicly available for verification, extension, citation, or reuse.

As part of MORF, we are assembling an open-source library of ready-to-use Docker containers to replicate experiments conducted on MORF to serve as shared baseline implementations. These containers can be loaded with a single line of code, allowing the research community to replicate, fork, interrogate, modify, and extend the results of an experiment. For example, the experiment in \cite{Gardner2018-ad} can be loaded by running \texttt{docker pull themorf/morf-public:fy2015-replication} in the terminal of any computer with Docker installed.

Building Docker containers requires only a single Dockerfile (akin to a makefile) which contains instructions for building the environment. This imposes minimal additional burden on researchers, but achieves a considerable increase in reproducibility over merely sharing code. Users can also generate Docker images by manually building a Docker environment (without a Dockerfile), or by using a tool such as Jupyter's \texttt{repo2docker}, which generates Docker images from public Github repositories\footnote{\url{https://github.com/jupyter/repo2docker}}.

MORF's combination of containerization and controller scripts allow the user to manage low-level experimental details (operating system and software dependencies, feature engineering methods, and statistical modeling) by constructing the Docker container which is submitted to MORF for execution, while MORF manages high-level implementation details (parallelization, data wrangling, caching of results) using the instructions provided in the controller script. Most importantly, the use of Docker containers ensures end-to-end reproducibility and enables sharing of the containerized experiment. 

\subsubsection{Other Services and Platform Architecture}

While Docker containers allow jobs submitted to the MORF platform to use any combination of programming languages, software, or analytical tools that can be installed on a linux system, MORF itself is written in Python and the platform utilizes a variety of Amazon Web Services tools, including S3, Lambda, Simple Queueing Service (SQS), Simple Email Service (SES), and others. The platform is built on top of the Flask web framework. MORF utilizes Zenodo to assign unique DOIs to every experimental artifact (configuration files and controller scripts), and Docker Hub to create a public archive of every Docker image executed on the platform.

\subsection{Platform Use and Software API}

\begin{figure}
    \centering
    \includegraphics[width = \columnwidth]{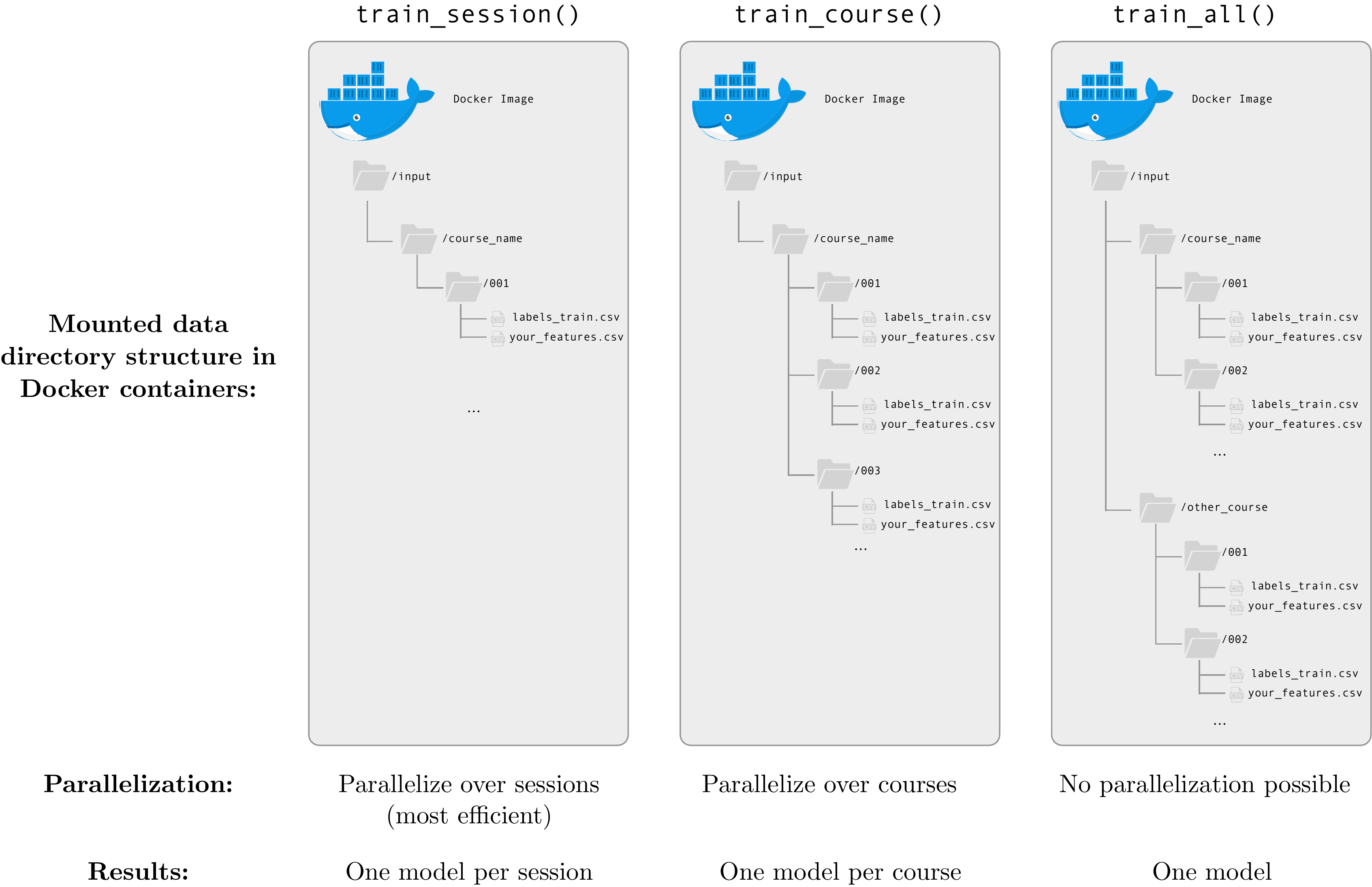}
    \caption{Example of MOOC data mounted to Docker images in MORF based on the use of varying API functions in a predictive modeling experiment. This shows how users control workflows using the Python API, while having the flexibility to process and model raw data using any underlying software within their Docker images.}
    \label{fig:train-data-mount}
\end{figure}

MORF includes a simple, minimal software API which is used to write controller scripts that control job execution on the platform (see Listing 1 for an example). The Python API allows users to provide a simple execution ``recipe'' for the MORF platform to execute their experiment specifying the complete end-to-end pipeline from raw data to model evaluation: \texttt{extract} features from raw data; \texttt{train} and \texttt{test} machine learning models (predictive modeling experiments only), and \texttt{evaluate} the experimental results. For example, after extracting the desired features, a predictive modeling experiment could train individual models for every session of a course by using \texttt{train\_session()} in their controller script; train one model per course using the data from all sessions by using \texttt{train\_course()}; or train a single monolithic model using all data from every session of every course by using \texttt{train\_all()}. An example of how MORF translates the various \texttt{train} functions into different experimental workflows, by mounting different data into the Docker environment, is shown in Figure \ref{fig:train-data-mount}. Note that the API functions used in the controller script are also used for MORF to manage many other low-level tasks, such as caching of intermediate results (e.g. trained models) and parallelization to ensure optimal performance. Similar functions exist for feature extraction, model testing, and evaluation, as well as for special cases (e.g. \texttt{fork}ing features extracted in a previous experiment for a new experiment).

The MORF API is written in Python, is available on the Python Package Index (PyPi)\footnote{\url{https://pypi.org/project/morf-api/}}, and is \texttt{pip}-installable. Development releases are available from Github.\footnote{\url{https://github.com/educational-technology-collective/morf}} It is important to note that while the MORF API is written in Python, this is simply the language that is required for controller scripts which MORF interprets to control job execution; users' code inside their Docker containers can be written using any language(s) or software that can be installed in the linux-based Docker container.

\subsection{Replication and Analysis Functionality}

\subsubsection{Predictive Modeling}

A critical area of MOOC research to date has been the construction and analysis of predictive models of student success \cite{Gardner2018-lp}. Such models have the potential to drive personalized learner supports or platform modalities \cite{Whitehill2015-ap}, adaptive learning pathways \cite{Pardos2017-lh}, contribute to learning theory or support data understanding (such as about demographic differences in dropout or achievement) \cite{Kizilcec2015-gu}, or ``early warning'' systems designed to alert instructors of struggling students.

One primary thread of this research has been predicting whether a student is likely to dropout of, or fail to complete, a course. MORF currently supports two dropout prediction tasks: binary classification (dropout/no dropout) and regression (the week a student will drop out of a course, up to the final week). Predictive modeling experiments on MORF follow a standard end-to-end supervised learning workflow: feature extraction from raw data; model training; model testing; and model evaluation (whereby performance is analyzed or, optionally, evaluated using statistical tests). In order to perform a predictive modeling experiment using MORF, users specify their Docker image should respond when executed in each of these ``modes'' with the requisite data mounted within the image. Predictive modeling experiments conducted on MORF to date include \cite{Gardner2018-ad, Gardner2018-wo}.

\subsubsection{Production Rule Analysis}

A key design principle of MORF has been to support high-quality replication research, and the generation of other novel findings, even for researchers not interested in (or technically capable of) predictive modeling. As such, MORF supports a second mode of evaluation, known as \textit{production rule} analysis. Production rules are if-then rules, coded in the form of ``If a student who is $<$attribute$>$ does $<$operator$>$, then $<$outcome$>$'' \cite{Andres2016-ju}. MORF uses the JESS expert system language to code these production rules. In order to perform a production rule experiment using MORF, users provide a text file containing the production rule of interest. A set of attributes, operators, and outcomes are available in the MORF platform. Currently, the available attributes, operators, and outcomes are based on prior findings which we sought to replicate in the MORF platform, although future versions of the platform will allow users to specify their own definitions for each. Production rule experiments conducted on MORF to date include \cite{Andres2016-ju, Andres2018-mt}.

\subsection{Platform Data}

\subsubsection{Available MOOC Data}

To date, the privacy restrictions protecting MOOC learner data have hindered researcher access to large, diverse MOOC datasets. This has led to many studies evaluating only a small number of courses: a recent survey indicated that 70\% of prior predictive modeling experiments in MOOCs used datasets consisting of five or fewer unique courses \cite{Gardner2018-lp}. As discussed above, larger, more diverse and representative datasets are critical for a nascent field such as MOOC research, which is still developing consensus on many research questions. 

The MORF platform seeks to resolve this issue, and makes a large, diverse dataset available for analysis of platform users. Currently, MORF contains the complete raw data exports from 209 sessions of 77 unique MOOCs offered on the Coursera platform, representing nearly 30 million unique interactions on courses offered across two of Coursera's four founding partner institutions (the University of Michigan and the University of Pennsylvania). Descriptive statistics for the data currently available in MORF are shown in Table \ref{tab:morf-data}. The provision of this data for analysis to MORF users represents a considerable advancement of the field.

\begin{table}[]
\centering
\caption{Data available in the MORF Platform. This allows all experiments conducted on MORF to utilize massive, diverse datasets considerably larger than almost all other MOOC prediction research to date \cite{Gardner2018-lp}.}\label{tab:morf-data}
\begin{tabular}{ll}
\hline 
\textbf{Metric}                 & \textbf{Value}      \\ \hline 
Total MOOC Sessions    & 209        \\
Unique Courses         & 77        \\
Active Students        & 986,420    \\
Interactions           & 29,416,369 \\
Discussion Forum Posts & 1,695,297  \\
Assignments            & 231,066   \\ \hline 
\end{tabular}
\end{table}

\begin{figure}
\centering
\begin{subfigure}[b]{\columnwidth}
   \includegraphics[width=\columnwidth]{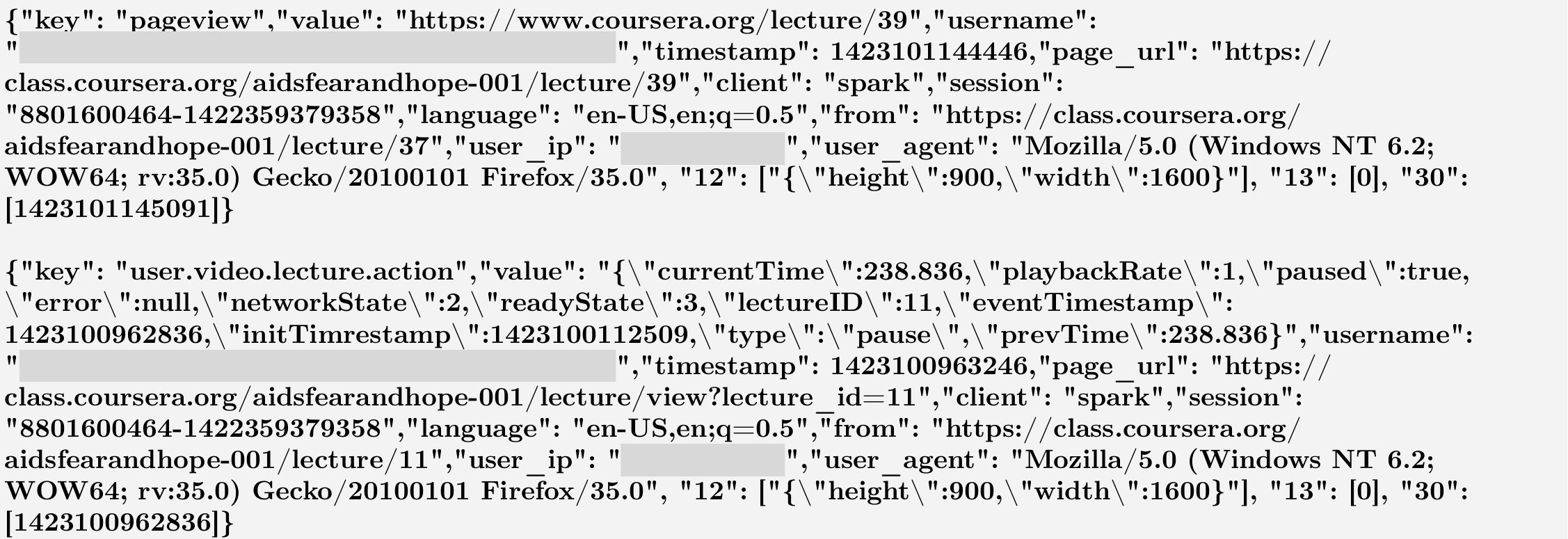}
   \caption{}
   \label{fig:clickstream-data} 
\end{subfigure}

\begin{subfigure}[b]{\columnwidth}
   \includegraphics[width=0.75\linewidth]{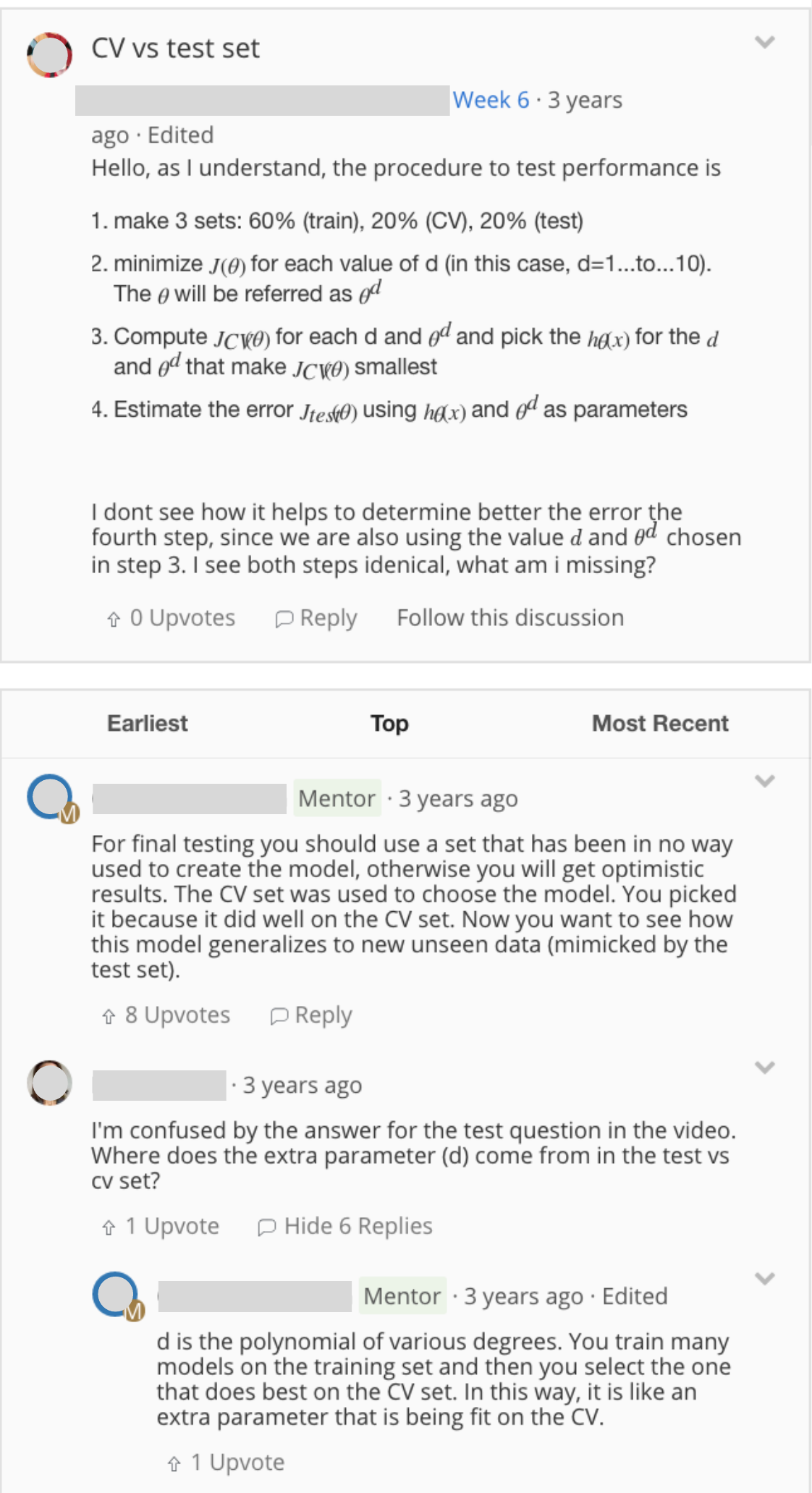}
   \caption{}
   \label{fig:forum-data}
\end{subfigure}

\caption[Data types]{Data types available for MOOC research in MORF. In existing solutions, these data sources are often highly redacted, anonymized, or simply unavailable. (a) Example clickstream log entries. (b) Example threaded forum posts. Personally-identifying information has been redacted.}
\end{figure}

For all of the courses in MORF, the complete, raw exports consisting of all available course, user, and interaction data are available. In the case of the Coursera data currently made available by MORF, this consists of seven exports for every course \cite{Coursera2013-jh}: multiple MySQL database dumps, CSV files, compressed clickstream server logs, and HTML files. This includes the complete clickstream server log for every user interaction with course content (example shown in Figure \ref{fig:clickstream-data}), discussion forum posts and associated metadata (time, thread ID, user ID, upvotes/downvotes received, etc.; shown in Figure \ref{fig:forum-data}), learner-provided responses to a course demographic survey, assignment submissions and grades with associated metadata, and course metadata (such as video titles, lengths and subtitle text; assignment information and due dates; etc.). Anonymizing such data, as many existing MOOC research solutions do, often obscures critical details, such as users' IP address (which can be used to estimate a users' location or socioeconomic status), the text of students' forum posts, and logfile entries which could be used to identify individual users. Using these rich datasets in raw and unredacted form presents an opportunity for researchers to investigate a highly diverse space of hypotheses, conduct rich feature engineering to build diverse predictive models, and replicate a wide variety of findings across many modes of inquiry.

% \begin{figure*}
% \centering
% \begin{subfigure}{.5\textwidth}
%   \centering
%   \includegraphics[width=4in]{img/clickstream-sample.png}
%   \caption{A subfigure}
%   \label{fig:sub1}
% \end{subfigure}%
% \begin{subfigure}{.5\textwidth}
%   \centering
%   \includegraphics[width=2.5in]{img/forum_thread.png}
%   \caption{A subfigure}
%   \label{fig:sub2}
% \end{subfigure}
% \caption{A figure with two subfigures}
% \label{fig:test}
% \end{figure*}

% another subfigure example
% \begin{figure}
% \centering
% \begin{minipage}{.5\textwidth}
%   \centering
%   \includegraphics[width=.4\linewidth]{image1}
%   \captionof{figure}{A figure}
%   \label{fig:test1}
% \end{minipage}%
% \begin{minipage}{.5\textwidth}
%   \centering
%   \includegraphics[width=.4\linewidth]{image1}
%   \captionof{figure}{Another figure}
%   \label{fig:test2}
% \end{minipage}
% \end{figure}

\subsubsection{Execute-Against Access for Secure Big Data Research}\label{sec:execute-against}
In order to provide direct and unredacted access to the rich, raw big data exports of MOOC platforms in a way that also accommodates the data regulations restricting the sharing of MOOC learner data (FERPA, IRB Common Rule, GDPR), we provide ``execute-against'' access to the platform data exports: users can run analyses against MORF's data, but cannot download or directly access the data. Instead, their experiments are executed against MORF's data within a sandboxed, networked-restricted computing environment. Then, a summary of the results are returned to the user. For predictive modeling jobs, these results are 8 measures of prediction performance (AUC, Cohen's $\kappa$, F1, etc.) on each course or session the job was executed on. For production rule analyses, statistical testing results are returned.

Most MOOCs are generated by a small number of platforms (e.g. Coursera, edX, FutureLearn), and all courses from a given platform use publicly-documented data schemas (e.g. \cite{Coursera2013-jh}). Thus, users can develop experiments using their own data from a given platform -- or even the public documentation -- and then submit these experiments for MORF to execute against \textit{any} other course from that platform. This enables MORF to provide an interface to its large data repository, without sharing the data itself, by utilizing the consistent and public schema of MOOC datasets. These shared public data schemas also ensure that existing experiments in MORF can be replicated against new data (from the same MOOC platform) as it becomes available. Additionally, while only descriptive summary results are returned to users, MORF persists all extracted features, trained models, and predictions obtained in the course of an experiment for complete reproducibility, which also allows for trusted users to obtain access to intermediate experimental results.

While the term ``execute-against'' access is novel, this framework of accepting analytical submissions which are executed against protected data has been used, for example, with healthcare data in the Critical Care Health Informatics Collaborative (CCHIC) \cite{Harris2018-vz} and with copyrighted music data in the Networked Environment for Music Analysis (NEMA) \cite{West2010-fy}. 

\subsection{MORF Resolves Existing Barriers to Replication in Educational Big Data Research}
MORF uses containerization to resolve many of the experimental challenges described in Section \ref{sec:barriers}. The Docker containers submitted to MORF fully encapsulate the code, software dependencies, and execution environment required of an experiment in a single file, ensuring end-to-end reproducibility and enabling sharing of the containerized experiment. These containers are automatically stored in a public repository and assigned unique, persistent identifiers. They can be loaded with a single line of code, allowing the research community to replicate, fork, interrogate, modify, and extend the results of any experiment in MORF.

\begin{figure}
    \centering
    \includegraphics[width = \columnwidth]{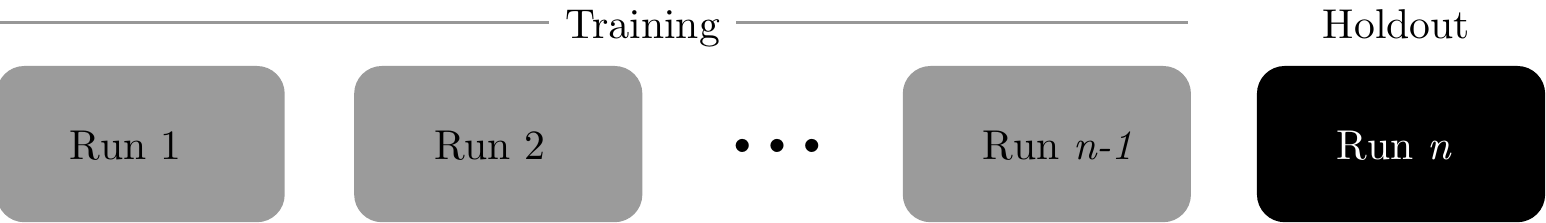}
    \caption{Default future-course prediction architecture used in MORF. This avoids potential inferential issues caused by using cross-validated performance to estimate model performance on future course sessions, by instead estimating that performance directly.}
    \label{fig:my_label}
\end{figure}

In order to address concerns with methodological and inferential reproducibility, MORF provides sensible default procedures for many tasks, such as model evaluation, as well as simple statistical testing procedures. For example, MORF avoids the use of cross-validation for model evaluation: the prediction tasks to which most MOOC models aspire are prediction of \textit{future} student performance (i.e., in an ongoing course where the true labels -- such as whether a student will drop out -- are unknown at the time of prediction). As such, using cross-validation within a MOOC session, when the outcome of interest is accuracy on a \textit{future} MOOC session, provides an unrealistic and potentially misleading estimate of model performance. Prior work has demonstrated that cross-validation provides biased estimates of independent generalization performance \cite{Varma2006-by}, and in the MOOC domain, that cross-validation can produce biased estimates of classification performance on a future (unseen) course \cite{Gardner2018-ad, Whitehill2017-tt}. Holding out a single session of every course requires a large data repository (multiple sessions of every MOOC). The rarity of such methods in prior work \cite{Gardner2018-xm} is likely due to the scarcity of MOOC data, an issue MORF resolves. Adopting more effective model evaluation techniques by default requires no additional work for MORF users, and ensures that work produced on the MORF platform follows effective model evaluation procedures. MORF also provides support for statistical model evaluation to provide for more reliable inference than comparing machine learning experiments without statistical testing, as is commonly performed \cite{Gardner2018-xm}.

MORF achieves data reproducibility while also meeting data privacy restrictions by providing ``execute-against'' access to underlying data, described in Section \ref{sec:execute-against}.

\begin{table}[]
\begin{tabular}{lllll}
 & \rotatebox{90}{MORF} & \rotatebox{90}{LearnSphere \cite{Koedinger2010-yg}*} & \rotatebox{90}{CodaLab} & \rotatebox{90}{CCHIC \cite{Harris2018-vz}} \\ \hline 
\textbf{Data} &  &  &  &  \\
\hspace{0.5cm}Raw (non-anonymized) data available & \yes & \some & NA & \some \\
\hspace{0.5cm}Complete MOOC exports available & \yes & \some & NA & NA \\
\hspace{0.5cm}Total MOOC Sessions Available & 209 & 151 & NA & NA \\
\hspace{0.5cm}Supports ``execute-against'' data & \yes & \no & \no & \yes \\ \hline 
\textbf{Platform} &  &  &  &  \\
\hspace{0.5cm}Supports custom code & \yes & \yes & \yes & \no \\
\hspace{0.5cm}Supports R, Python, Java, C++ & \yes & \yes & \yes & \some \\
\hspace{0.5cm}Command-line submission/execution & \yes & \no & \yes & \no \\
\hspace{0.5cm}Open-Source & \yes & \yes & \yes & \yes \\
\hspace{0.5cm}GUI Available & \no & \yes & \yes & \no \\ \hline 
\textbf{Reproducibility} &  &  &  &  \\
\hspace{0.5cm}Uses containerization & \yes & \no & \yes & \no \\
\hspace{0.5cm}Accepts submission as container & \yes & \no & \no & \no \\
\hspace{0.5cm}End-to-end execution in platform & \yes & \some & \yes & \no \\
\hspace{0.5cm}Assigns DOI to each job & \yes & \no & \no & \no \\
\hspace{0.5cm}Analyses are forkable & \yes & \some & \yes & \some \\
\hspace{0.5cm}Integrated w/public image repository & \yes & \no & \no & \no \\ \hline 
\textbf{Computation} &  &  &  &  \\
\hspace{0.5cm}Provides computational resources & \yes & \yes & \yes & \no \\
\hspace{0.5cm}Native Parallelization Implemented & \yes & \no & \no & \no \\
% \hspace{0.5cm}Number of CPUs Available & 64 & ? & 6\dagger & NA \\ \hline 
\end{tabular}
\caption{Comparison to existing solutions with similar goals, including those within the same domain (Learnsphere + Datastage), in a related domain (CCHIC), and domain-agnostic tools (Codalab). \yes = Supported; \some = Partially Supported. *: statistics shown for use with DataStage MOOC data repository. $\dagger$: users can provide their own additional worker nodes.}
\end{table}

\section{Discussion}\label{sec:discussion}

\subsection{Implications for Big Data Research}

In addition to jointly addressing several challenges to reproducible and replication research within the field of education, MORF's architecture, workflow, and initial research results have implications for the broader big data community. MORF addresses a set of problems faced by big data researchers across many domains. This includes experimental reproducibility as big data research in many fields uses increasingly complex computational models; methodological and inferential reproducibility as big data research enables problematic statistical practices such as massively multiple testing via testing thousands or millions of hypotheses in a single experiment; and data reproducibility as available data become massively multimodal (many different formats), measure increasingly private or restricted aspects of users' behavior and identity, and cannot be easily anonymized. 

MORF demonstrates that big data research in the face of such challenges is possible, and suggests a general blueprint for conducting a variety of research tasks while addressing these issues. MORF's API code is fully open-source, as is most of the code supporting the MORF platform (except where doing so presents a security risk). The MORF framework is domain-agnostic, and can support generic workflows for supervised learning and production rule analyses in any domain which works with complex, multiformat data which cannot be easily anonymized  (e.g. sensitive medical data, copyrighted media, computational nuclear physics). As noted in Section \ref{sec:execute-against} above, systems using execute-against access have been successfully deployed in the fields of healthcare and music information retrieval.

\subsection{Beyond Verification: The Benefits of Replication}

Much prior work on reproducibility has focused on verifying that published results can be reproduced. However, end-to-end reproducible machine learning frameworks, such as MORF, provide benefits beyond mere verification, including:

\textbf{Gold standard benchmarking:} open replication platforms allow for the comparison of results which were previously not comparable, having been conducted on different data. The use of such benchmarking datasets has contributed to the rapid advance of fields such as computer vision (e.g. MNIST, IMAGENET), natural language processing (Penn Tree Bank, Brown corpus), and computational neuroscience (openFMRI). These datasets have been particularly impactful in fields where it is difficult or expensive to collect, label, or share data (as is the case with MOOC data, due to legal restrictions on sharing and access). These help to evaluate the state of the art by providing a common performance reference which is currently missing in many fields.

\textbf{Shared baseline implementations:} We noted above that variability in so-called ``baseline'' or reference implementations of prior work has contributed to concerns about reproducibility in the field \cite{Henderson2018-qd}. By providing fully-executable versions of existing experiments, MORF ameliorates these issues, allowing for all future work to compare to the exact previous implementation of a baseline method.

\textbf{Forkability:} containerization produces a ready-made executable which fully encompasses the code and execution environment of an experiment. These can be readily shared and ``forked'' across the scientific community, much in the same way code is ``forked'' from a git repository. This allows researchers to build off of others' work by modifying part or all of an end-to-end pipeline (for example, by experimenting with different statistical algorithms but using the same feature set as a previous experiment) within the same software ecosystem. 

\textbf{Generalizability analysis:} Each successive replication of an experiment provides information about its generalizability. Evaluating the generalizability of experiments has been a challenge in MOOC research to date, where studies conducted on single-course, restricted, and often homogenous datasets tend to dominate the literature. When containerized implementations are available, replicating these analyses on new data -- even data which are not publicly available but share the schema of the original data -- becomes as straightforward as running the containerized experiment against new data.

\textbf{Sensitivity Analysis:} This technique, used widely in Bayesian analysis, evaluates how changes to the underlying assumptions or hyperparameters affect experimental results. Such an evaluation can provide useful information about a model's robustness and potential to generalize to new data. Without being able to fully reproduce a model on the original data, sensitivity analyses of published results are not possible. In MORF, such analyses can be conducted by forking and modifying the containerized version of the original experiment, then re-executing it against the same data. This process also enables so-called \textit{ablation analyses}, wherein individual components are removed from a model to observe their contribution to the results, as well as \textit{slicing analyses}, where analysis of performance across different subgroups (e.g. demographic groups) is explored \cite{Sculley2018-md}.

\textbf{Full Pipeline Evaluation:} Each stage of an end-to-end machine learning experiment (feature extraction, algorithm selection, model training, model evaluation) can be done in many different ways. Each stage also affects the others (for example, some algorithms might perform best with large feature spaces; others might perform poorly with many correlated features). However, current research usually evaluates only one or two components of this pipeline (e.g. training several algorithms and tuning their hyperparameters on a fixed feature set). Not only are the remaining stages often described in poor detail or not at all \cite{Gundersen2017-zl}; such work also leaves future researchers unable to evaluate the synergy between different aspects of the end-to-end pipeline in a published experiment (for example, exploring whether an algorithm's performance improves with a different feature set). MORF fully encapsulates this end-to-end pipeline for a given experiment and it makes it available for modification to any other researcher.

\textbf{Meta-Analysis:} While meta-analyses are common in fields with robust empirical research bases, such analyses have been less common in the field of big data, which has an emphasis on novelty. The open availability of executable machine learning experiments affords detailed meta-analyses by providing complete results of all modeling stages for meta-analysis.  MORF has already been used for meta-analysis \cite{Andres2018-mt}.

\section{Conclusion}

Replication is important to all research, including big data research in education. We have constructed MORF, a system to facilitate such analysis. MORF has been used to conduct several replication studies \cite{Gardner2018-ad, Gardner2018-wo, Andres2016-ju, Andres2018-mt}, each of which demonstrated that the initial findings did not entirely replicate. While we encourage interested institutions to partner with us to make their data available within MORF, users or institutions can also create their own platform instances by using the open-source code and API. We see immediate future opportunities to broaden our work by (a) reducing the need for researchers to understand environments such as Docker, by integrating front-end calls to this tool within development environments such as Jupyter; (b) providing light weight web front-ends for this system, allowing users to explore this large data through web visualizations; and (c) increasing the size and scope of the project by involving more institutions and a broader array of educational data (e.g. non-MOOC learning management system data).

% \section*{Acknowledgment}

\bibliographystyle{IEEEtran}
\bibliography{MORF_IEEEBD}

\end{document}